\documentclass[10pt,conference]{IEEEtran}
\usepackage[T1]{fontenc}
\usepackage{cite}
\usepackage{amsmath,amssymb,amsfonts}
\usepackage{graphicx}
\usepackage{textcomp}
\usepackage{xcolor}
\usepackage[hyphens]{url}
\usepackage{hyperref}
\hypersetup{hidelinks}
\usepackage{multirow}
\usepackage{soul}
\usepackage{makecell}
\usepackage{threeparttable}

\usepackage{algorithm}
\usepackage{algpseudocode}
\usepackage{textcomp}
\usepackage{inconsolata}

\usepackage{siunitx}
\usepackage{xspace}

\newcommand{\us}[1]{\SI{#1}{\micro\second}}

\newcommand{\ndp}{KARAT}  
\newcommand{\gpnm}{KARAT}  
\newcommand{\simt}{\raise.30ex\hbox{$\scriptstyle\sim$}}

\newcommand{\ub}[1]{\textmu B#1}

\title{Heterogeneous LLM Serving with  \\ General-Purpose Processing-Near-Memory for\\
Retrieval-Based Sparse Attention}

\author{
\IEEEauthorblockN{Hyungkyu Ham, Junhyeong Bae, Seungheon Lee, Myeongjae Jeon, and Gwangsun Kim}
\IEEEauthorblockA{Pohang University of Science and Technology (POSTECH)}
}

\begin{document}

\bstctlcite{IEEEexample:BSTcontrol}

\maketitle
\thispagestyle{plain}
\pagestyle{plain}

\begin{abstract}

This paper presents a heterogeneous decode-phase serving system that relocates
the KV cache out of GPU memory, motivated by the retrieval-based sparse
attention that recent frontier LLMs adopt to serve million-token contexts.
It partitions a decode step by operation type: GPU nodes hold the model weights
and execute the projections and MoE layers, while processing-near-memory (PNM)
nodes hold the KV cache and index keys and execute every operation that reads
them.
We first show that the assumptions behind prior PIM and PNM designs no longer
hold for these operations, and derive four design requirements for such a node.
From these requirements, we propose \emph{KARAT} (KV-cache-resident Accelerator
for Retrieval-based ATtention), a general-purpose PNM design that is the design
point meeting all four.
A KARAT device combines large LPDDR capacity with general-purpose compute
sized for the retrieval indexer, serving an operational intensity beyond what
PIM/PNM designs built for low-intensity GEMV target while accommodating diverse
sparse attention algorithms that fixed-function units cannot support as they
evolve.
To reduce pipeline bubbles as the two device types alternate between
micro-batches, we further propose
\emph{opportunistic, fine-grained micro-batch scheduling} (OFMS), which hides
expert all-to-all behind the other micro-batch's GEMMs, and
\emph{context-length-aware micro-batch rebalancing} (CMR), which equalizes their
token counts despite the variance in context length.
Across three state-of-the-art models and real agentic traces, our proposed system
improves throughput per TDP under a service-level objective by 2.09--6.13$\times$
over a GPU-only baseline and runs training-free sparse attention methods
with 1.36--3.21$\times$ improvements.

\end{abstract}

\section{Introduction}

Large language models (LLMs) are now deployed across many
domains~\cite{chatgpt, gemini_2.5}, and increasingly serve agentic workloads, in
which the model itself drives multi-turn generation
loops~\cite{claude-opus-5, ReAct, Toolformer} and tool outputs and intermediate
reasoning accumulate over a session.
Real agentic traces reach 198K--791K tokens at the 99th percentile, with
94--97\% reused on the following turn (\S\ref{sec:bg_longcontext}).
Prefill extends each context by only a few hundred tokens per turn, whereas
every decoding step must compute attention over the entire context, so the
decode pool---which prefill--decode disaggregation now provisions
separately~\cite{splitwise, distserve, mooncake, nvidia_dynamo}---must both hold the context of
every concurrently served request in memory and attend over all of them at every
step.
Capacity and bandwidth therefore become the bottleneck, and model architectures
have evolved to relieve both: MLA (multi-head latent attention) compresses the
KV cache into a low-rank latent~\cite{deepseek-r1, deepseek_v3, glm-5.2}, and
sparse attention reads only a subset of it per
step~\cite{quest, magicpig, deepseek-v3.2, minimax-m3}.

\begin{figure}
    \centering
    \includegraphics[width=1.0\linewidth]{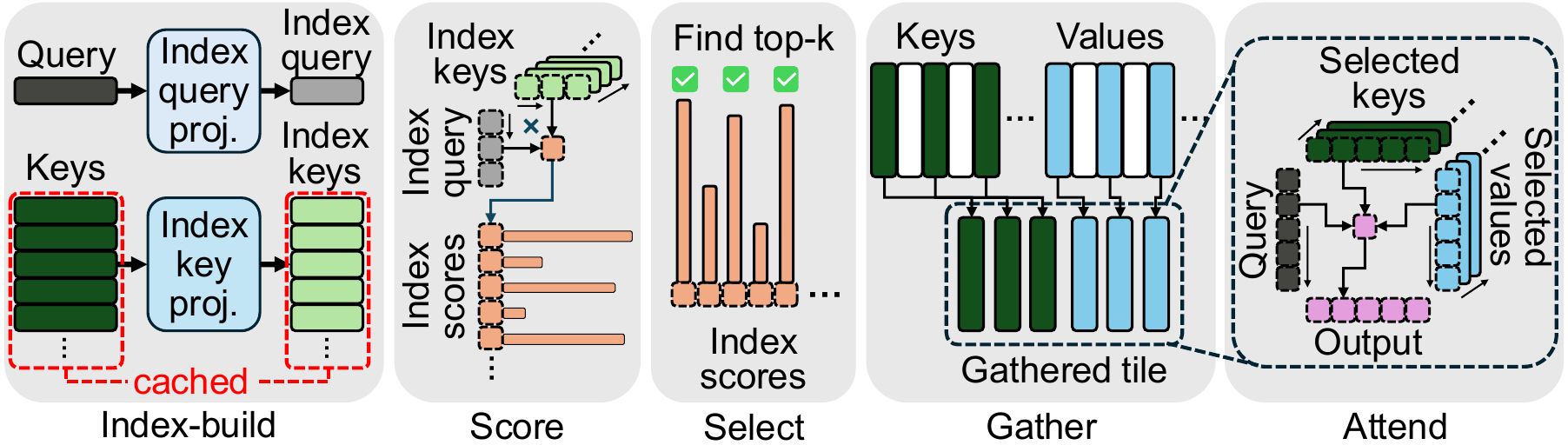}
    \vspace{-.15in}
    \caption{Stages of retrieval-based sparse attention in state-of-the-art LLMs.}
    \label{fig:retrieval-based-sparse-attention}
\end{figure}

\begin{figure}
    \centering
    \includegraphics[width=0.7\linewidth]{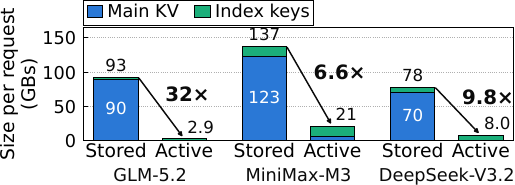}
    \caption{Stored and activated KV cache and index keys at 1M context for the
  three evaluated models. Retrieval-based sparse attention reduces the activated
  data that each decoding step reads, while the full KV cache and index keys  must still be stored.}
    \label{fig:kv_cache_footprint}
\end{figure}

In particular, \emph{retrieval-based sparse attention} has been adopted by
frontier LLMs as part of the model architecture.
It layers on top of a KV-reducing attention form rather than replacing one:
over MLA in GLM-5.2~\cite{glm-5.2} and DeepSeek-V3.2~\cite{deepseek-v3.2}, and
over grouped-query attention (GQA) in MiniMax-M3~\cite{minimax-m3}.
It stores a small index key for each of the $L$ context tokens, and at every decoding step scores
the query against all index keys, selects the top-$k$ tokens, and attends only
to them (Fig.~\ref{fig:retrieval-based-sparse-attention}).
This reduces attention FLOPs and KV reads from $O(L)$ to $O(k)$---6.6--32$\times$
less KV-cache traffic at a million-token context
(Fig.~\ref{fig:kv_cache_footprint}).
Unlike eviction-based compression~\cite{h2o, snapkv}, which bounds the footprint
by discarding KV pairs at the cost of irrecoverable context loss, these
methods retain every pair and add per-token index keys, reducing memory traffic
but not the memory footprint.
A single million-token request holds 78--137\,GB of KV cache and index keys,
whereas a recent eight-GPU node~\cite{dgx_h200, dgx_b200} provides 1.1--1.4\,TB
of HBM that must also hold the hundreds of billions of parameters of mixture-of-experts (MoE) 
models~\cite{glm-5.2, deepseek-v3.2, minimax-m3}.
Few requests therefore fit in GPU memory at once, and the dense FC and MoE
layers spend $58$--$74\%$ of each decoding step on weight reads that a small
batch cannot amortize (\S\ref{sec:eval}).
Thus, throughput is limited by the attainable batch size rather than by the
available compute.

Moreover, indexing and attention are not purely bandwidth-bound.
Depending on the indexer's score function and the attention form that consumes
its output---MLA in particular---a substantial part of them is 
compute-intensive.
They therefore demand capacity, bandwidth, and compute at once.
The MoE layers, in contrast, are compute-bound only at a large batch size,
determined by the capacity left for the KV cache.

Two lines of prior work place the KV cache outside GPU memory, but neither
addresses all three together.
Offloading it to host DRAM~\cite{sglang, hisparse, lmcache, mooncake} or to
CXL-attached memory~\cite{tract, sac, octopus} expands capacity without
relocating the computation, 
so the $O(L)$ index scan over all index keys and the
gather of the selected tokens still cross a bandwidth-limited link.
Demoting the KV caches of \emph{idle} sessions to a lower memory
tier~\cite{mooncake, strata} is complementary but does not address the
\emph{active} sessions.
Recent processing-in-memory (PIM) and processing-near-memory (PNM) designs instead
relocate the computation next to the memory, but assume both the low operational
intensity (OI) and the fixed form of dense multi-head attention (MHA) and
GQA~\cite{attacc, neupims, cent, duplex, m2ndp}.
Under these assumptions, they keep their compute engines small and implement
them as specialized PNM (SPNM) units tailored to those attention
forms~\cite{cxl-llm-pact25, starc}.
Retrieval-based sparse attention methods, however, vary in the score function,
the selection granularity, and the cadence at which the indexer runs, and new
variants continue to emerge.
An architecture coupled to any single method can therefore lose its
advantage~\cite{mtia_v1} and impede algorithmic progress.

Because datacenters provision serving systems under a fixed power budget, these
observations motivate a device that holds the KV cache and index keys at high
capacity and bandwidth per watt, with a compute engine placed next to them.
The engine must be general-purpose, since specialized units cover only a fixed
set of algorithms, and sized to the moderate OI of 4--128 across the algorithms
in use (\S\ref{sec:bg_sparse}).
The GPUs, in contrast, must retain high compute throughput per watt, since the
MoE layers become compute-bound at large batch sizes, which the added
capacity can now permit.

To meet these requirements, we propose a scalable heterogeneous decode-phase
serving system that partitions a decode step by operation type: GPU nodes hold
the model weights and execute the projections and MoE layers, while our proposed \gpnm{}
(\emph{\textbf{K}V-cache-resident \textbf{A}ccelerator for
\textbf{R}etrieval-based \textbf{AT}tention}) nodes hold the KV cache and index
keys and execute all operations that read them.
A \gpnm{} device combines LPDDR memory, which provides high memory capacity per
watt, with general-purpose PNM (GPNM) cores that sustain the OI of the target
operations while supporting diverse retrieval-based sparse attention algorithms.
Multiple (e.g., 32) such devices form a \gpnm{} node that matches the power and
aggregate memory bandwidth of a GPU node with an order of magnitude more memory
capacity.
Offloading the KV cache to \gpnm{} nodes dedicates GPU HBM to the model weights,
raising the global batch size at which the MoE layers become compute-bound under
the same power constraint.

A decode step interleaves GPU and \gpnm{} computation, so each batch is split
into two micro-batches that pipeline the two device types. The duration of each
stage, however, varies with the aggregate context length of a micro-batch, the
expert skewness within the MoE layers, and the layer-wise variation of the
model, leaving pipeline bubbles on whichever device is off the critical path. We
therefore propose two scheduling techniques, one for each side of the pipeline.

First, when GPU operations dominate, the GPUs still remains idle during the all-to-all
communication of the MoE layer. Prior micro-batch schedulers~\cite{neupims,
duplex, megascale-infer} alternate between compute-bound GPU kernels and
memory-bound PIM/PNM kernels in a static, coarse-grained manner, leaving this
idle time unexploited. Thus, we propose \emph{opportunistic, fine-grained
micro-batch scheduling} (OFMS), which overlaps the communication with the other
micro-batch's next GEMM whenever dependencies allow, and falls back otherwise.

Second, when \gpnm{} operations dominate, the bottleneck can instead come from micro-batch
imbalance: the wide variance in context length across requests can leave the two
micro-batches unbalanced in total tokens. Prior
approaches~\cite{neupims, megascale-infer, exegpt} balance them by request count and
overlook the aggregate context length, so the longer micro-batch delays the
synchronization at the end of the step. We therefore propose
\emph{context-length-aware micro-batch rebalancing} (CMR), which equalizes the
token count of the two micro-batches whenever a request finishes or is admitted.

On agentic multi-turn and long-context workloads, \ndp{} improves
SLO-constrained decode throughput per thermal design power (TDP) by
$2.09$--$6.13\times$ across three state-of-the-art models, relative to a
GPU-only baseline running the same model-native method under a P99
time-between-tokens (TBT) service-level objective (SLO) of
100\,ms~\cite{sarathi_serve, splitwise}.
It also runs training-free sparse attention methods without hardware change,
improving throughput per watt by $1.36$--$3.21\times$ over the GPU-only system.

To summarize, this work makes the following contributions:
\begin{itemize}
    \item \textbf{Characterization.} We show that the assumptions behind prior
      PIM and PNM designs no longer hold, and derive four design requirements
      for serving diverse retrieval-based sparse attention algorithms.
\item \textbf{General-purpose PNM.} We characterize the PNM design space for
      retrieval-based sparse attention and identify a design point that meets all
      four requirements: LPDDR chosen for capacity per watt, with
      general-purpose cores provisioned only to the intensity of the offloaded
      operations, which we refer to as \gpnm{}. Its GPNM cores support diverse
      retrieval-based sparse attention algorithms that specialized PNM cannot
      accommodate.
    \item \textbf{Micro-batch scheduling.} We propose \emph{opportunistic,
      fine-grained micro-batch scheduling} (OFMS), which interleaves GPU kernels
      across micro-batches to hide expert all-to-all communication.
    \item \textbf{Micro-batch rebalancing.} We propose \emph{context-length-aware
      micro-batch rebalancing} (CMR), which keeps the token counts of the
      micro-batches comparable despite the wide variance in context length.
\end{itemize}

\vspace{-.02in}
\section{Background and Motivation}
\vspace{-.03in}
\label{sec:background}

\newcommand{\obs}[1]{\textbf{(\textsf{#1})}\;}
\newcommand{\obsref}[1]{\textsf{\textbf{#1}}}

Throughout this section, we mark the observations that motivate our design in
italics as \obsref{O1}--\obsref{O5}.

\subsection{Long-context Agentic LLM Serving}
\label{sec:bg_longcontext}

\begin{figure}
    \centering
    \includegraphics[width=0.9\linewidth]{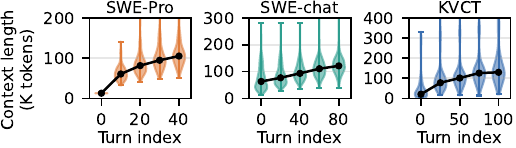}
\caption{Context length distribution of multi-turn agentic workloads~\cite{swe-chat, swepro, kvct}:
Claude Code traces with Claude Opus~\cite{claude-opus-4-6} (SWE-chat, KVCT)
and OpenAI~\cite{openai_models} coding-agent rollouts (SWE-bench Pro).
Each dot denotes the median context length at a given turn, and we plot turns
up to the 75th percentile (P75) of the turn-count distribution.}
    \label{fig:agentic_context}
    \vspace{-.05in}
\end{figure}

Agentic workloads drive a multi-turn loop in which tool outputs and intermediate
reasoning accumulate over a session~\cite{ReAct, Toolformer}.
Each turn resubmits the entire context and appends to it, so the context grows
monotonically (Fig.~\ref{fig:agentic_context}).
Serving systems exploit this by caching the shared prefix rather than
recomputing it~\cite{sglang, vllm}.
\obs{O1}\emph{Across three real agentic traces, context accumulates over turns
to million scale (p99 198K--791K, max ${\sim}1$M) with high dispersion across
requests (p50 ${\sim}$83--149K), and 94--97\% of a turn's prompt tokens hit the
prefix cache, leaving only the remainder to be prefilled.}

The two phases bear these costs asymmetrically.
Prefill is compute-bound and traverses only the few hundred uncached tokens
once.
Decode reuses the context at a far finer granularity: every token a turn
generates attends over the entire accumulated context, so the same KV cache is
read on every step and must stay close to the compute for the whole turn.
Its footprint therefore caps the batch size, leaving the tensor cores idle while
the memory is full.
Production stacks accordingly disaggregate the two phases into separately
provisioned pools~\cite{splitwise, distserve, mooncake}, and we focus on the
decode pool throughout, holding the prefill pool constant.

\subsection{Retrieval-Based Sparse Attention}
\label{sec:bg_sparse}

To serve million-token contexts, sparse attention selects at runtime only the
tokens most relevant to the current query, reducing attention FLOPs and KV reads
from $O(L)$ to $O(k)$.
\emph{Eviction-based} methods (e.g., StreamingLLM~\cite{streaming-llm}, H2O~\cite{h2o},
SnapKV~\cite{snapkv}) discard the rest, bounding memory at the cost of
irrecoverable context loss.
\emph{Retrieval-based} methods instead retain the full KV cache and retrieve a
fresh top-$k$ set at every step, avoiding such loss.
Among them, training-free approaches~\cite{quest, sparq, retroinfer} attach a
selection mechanism to a pretrained model, whereas \emph{model-native} designs
train the indexer jointly with the model and are now part of the architecture of
frontier LLMs~\cite{deepseek-v3.2, glm-5.2, minimax-m3}.
\obs{O2}\emph{Retrieval-based sparse attention cuts KV-cache read traffic by
6.6--32$\times$ at a 1M-token context (Fig.~\ref{fig:kv_cache_footprint}), yet
the full KV cache and its index keys must still be stored alongside MoE weights
of hundreds of billions of parameters, shifting the decode bottleneck from
memory bandwidth to capacity.}

Their operational forms are also diverse.
DeepSeek Sparse Attention (DSA)~\cite{deepseek-v3.2} scores each token as a weighted sum of ReLU-rectified
dot products over a few FP8 indexer heads sharing one key head, and takes an
exact global top-$k$ at every layer.
GLM-5.2 reuses that scoring function but evaluates it once every four
layers~\cite{glm-5.2}, while MiniMax-M3 scores 128-token blocks per GQA
group~\cite{minimax-m3}, making the gather contiguous but the selection coarse.
Training-free methods add page-level min/max scans~\cite{quest},
LSH~\cite{magicpig}, and clustering~\cite{starc}.

\begin{table}
\centering
\caption{Representative operational forms at each stage of retrieval-based sparse
attention.} 
\footnotesize
\begin{tabular}{l|p{6.2cm}}
\hline
Stage & Observed operational forms \\ \hline\hline
\textsc{Index-Build} & learned projection~\cite{deepseek-v3.2, glm-5.2}, cross-layer reuse~\cite{glm-5.2},
page min/max~\cite{quest}, online clustering~\cite{starc}, LSH bucketing~\cite{magicpig}, segmented clustering~\cite{retroinfer} \\ \hline
\textsc{Score} & weighted MH dot~\cite{deepseek-v3.2, glm-5.2}, block score~\cite{minimax-m3}, sign-selected dot~\cite{quest},
partial dot~\cite{sparq}, centroid GEMV~\cite{starc, retroinfer}, LSH collision count~\cite{magicpig} \\ \hline
\textsc{Select} & \emph{exact} global top-$k$~\cite{deepseek-v3.2, glm-5.2},
block top-$k$~\cite{minimax-m3}, page top-$k$~\cite{quest},
cluster top-$k$~\cite{starc,retroinfer}, LSH sampling~\cite{magicpig} \\ \hline
\textsc{Gather} & token~\cite{deepseek-v3.2, glm-5.2}, block~\cite{minimax-m3}, page~\cite{quest}, cluster~\cite{starc} \\ \hline
\textsc{Attend} & MLA~\cite{deepseek-v3.2, glm-5.2}, GQA~\cite{quest, minimax-m3, starc, magicpig, sparq} \\ \hline
\end{tabular}
\label{tab:op_diversity}
    \vspace{-.05in}
\end{table}

\begin{table}[h!]
  \centering
  \caption{Evaluated mechanism families and their kernel characteristics
  (measured serving configurations; byte sizes are per token, summed over all
  layers).}
  \setlength{\tabcolsep}{4pt}\scriptsize
  \begin{tabular}{l|c|c|c}
  \hline
  Model & \makecell{GLM-5.2\\(MLA+DSA)} & \makecell{MiniMax-M3\\(GQA+sparsity)} & \makecell{DeepSeek-V3.2\\(MLA+DSA)} \\ \hline\hline
  Layers (w/ indexer) & 78 (21) & 60 (57) & 61 (61) \\ \hline
  Indexer OI & ${\sim}$64 & ${\sim}$4 & ${\sim}$128 \\ \hline
  Indexer cadence & every 4th layers & nearly every layer & every layer \\ \hline
  \textsc{Select} granularity & token & block (128) & token \\ \hline
  \textsc{Attend} OI & ${\sim}$114 & ${\sim}$16 & ${\sim}$228 \\ \hline
  Top-$k$ (tokens) & 2048 & 128$\times$16 (+1 local) & 2048 \\ \hline
  KV cache (KB/tok.) & 89.9 & 122.9 & 70.3 \\ \hline
  Index key (KB/tok.) & 2.7 & 14.6 & 7.8 \\ \hline
  \end{tabular}
  \label{tab:models}
    \vspace{-.02in}
\end{table}

\smallskip\noindent{\bf Operational structure and diversity.}
Retrieval-based sparse attention proceeds in five stages:
\textsc{Index-Build}, \textsc{Score}, \textsc{Select}, \textsc{Gather}, and
\textsc{Attend}.
\textsc{Index-Build} generates the per-token \emph{index keys}; it is $O(1)$ per
decode step and can be fused with the QKV projection or run as a separate
kernel.
\textsc{Score} and \textsc{Select} together form the \emph{indexer scan}, which
reads the index keys of the entire context to choose the top-$k$ tokens.
\textsc{Gather} then collects the selected entries and \textsc{Attend} computes
attention over them;
as the latter directly consumes the former's output, the two can be fused
into a single kernel.

\begin{figure}
    \centering
    \includegraphics[width=.85\linewidth]{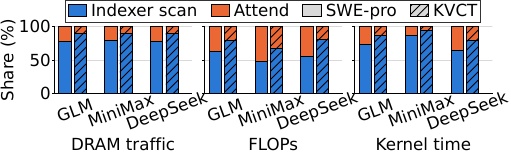}
    \caption{Decode cost of one layer that executes the retrieval indexer and sparse
  attention, at the average context lengths of the agentic traces
  (Table~\ref{tab:workloads}) with a decode batch of 256. Kernel time measured on
  H200.}
    \label{fig:attention_breakdown}
\end{figure}

Each stage takes a different operational form across models, methods, and
generations (Table~\ref{tab:op_diversity}), including the three frontier models
we evaluate (Table~\ref{tab:models}).
The scan, however, dominates regardless: it reads the entire context whereas
\textsc{Attend} touches only $O(k)$ tokens, so it accounts for most of the
memory traffic and the FLOPs of the attention layer for long contexts
(Fig.~\ref{fig:attention_breakdown}).
It therefore sets both demands at once, at an operational intensity (OI) of
4--128 FLOPs/byte that batching cannot shift: batching amortizes weight reads,
but the scan's FLOPs and bytes scale together.
\obs{O3}\emph{Retrieval-based sparse attention takes many operational forms, but
they share a common bottleneck: the indexer scan dominates the attention layer
rather than \textsc{Attend}. Its OI of 4--128 FLOPs/byte does not change with
batch size.}

\subsection{Expert and Data Parallelism for MoE LLM Serving}
\label{sec:bg_dpep}

To serve MoE LLMs with hundreds of billions of parameters, serving
systems~\cite{vllm, sglang, tensorrt_llm} combine data parallelism (DP), tensor
parallelism (TP), and expert parallelism (EP).
EP meets the memory capacity demand by distributing the experts across GPUs,
each executing only the tokens routed to it, while the dense weights are
replicated and attention runs in DP~\cite{megascale-infer}.
At every MoE layer, the ranks dispatch tokens to their experts and combine the
results after the expert GEMMs, both through all-to-all communication.

\begin{figure}
    \centering
    \includegraphics[width=0.8\linewidth]{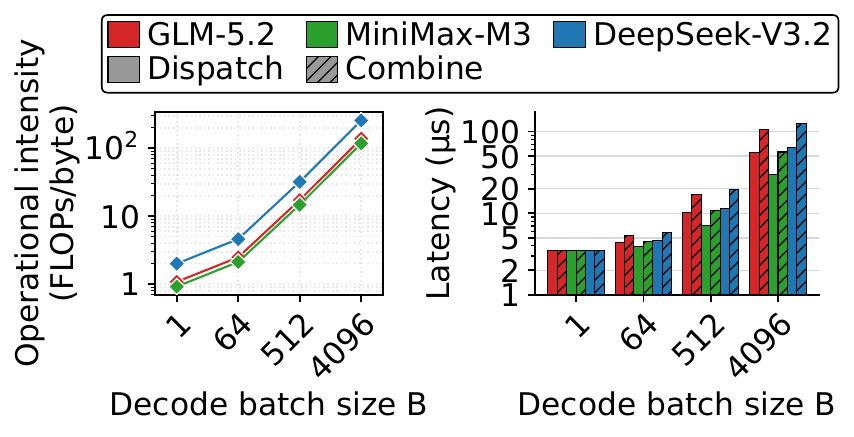}
    \\[-0.07in]
    {\scriptsize\raggedright\noindent%
        \hspace*{0.36\linewidth}(a)%
        \hspace{0.30\linewidth}(b)%
    \par}
    \caption{
  (a) OI of the MoE layer across batch sizes, assuming each expert's weights are
read from HBM and tokens are routed uniformly across experts.
(b) MoE communication latency across global batch sizes at $EP{=}64$.
        }
    \label{fig:roofline-OI}
\end{figure}

Each expert therefore processes only a fraction of the global batch: with 256
experts and eight selected per token, a global batch of 8192 yields an average
of 256 tokens per expert.
MoE efficiency thus hinges on the global batch, the layer's OI rising from
${\sim}1$ to over 100\,FLOPs/byte between $B{=}1$ and $B{=}4096$
(Fig.~\ref{fig:roofline-OI}(a)).
The all-to-all exchange, however, grows with the batch, reaching over
\us{100} at $B{=}4096$ under $EP{=}64$ (Fig.~\ref{fig:roofline-OI}(b)), and
skew in expert selection leaves the load uneven across ranks~\cite{deepseek_v3, faster-moe}.
The attainable batch size is nonetheless bounded by memory capacity, since the aggregate
KV cache of the batch must share the HBM with the model weights.
\obs{O4}\emph{The MoE layer reaches high efficiency only at a large global
batch, which is in turn limited by the HBM capacity remaining for the KV cache.
Throughput is therefore bounded by memory capacity rather than by compute.}

\subsection{Challenges in Prior Approaches}
\label{sec:bg_fail}

\smallskip\noindent{\bf Offloading the KV cache.}
Host- or CXL-attached memory~\cite{hisparse, sac, tract, beluga} expands
capacity but leaves the \emph{location of computation} unchanged, forcing a
choice between two bottlenecks.
Retaining the index keys in HBM avoids traversing the link with the $O(L)$ scan,
yet the keys grow with context length and with every concurrent request---they
are only 8--33$\times$ smaller than the KV cache per token
(Table~\ref{tab:models})---so HBM capacity still caps the global batch.
Offloading them instead places the scan's $O(L)$ traffic on a link an order of
magnitude slower than HBM, and that scan dominates KV-side time (\obsref{O3}).
Neither choice removes the bottleneck; it determines only which resource is
exhausted first.

\smallskip\noindent{\bf Offloading the KV cache and the computation.}
The computation must therefore move to the data.
The host CPU is an inadequate site, as its DRAM bandwidth falls an order of
magnitude below HBM's, whereas PNM and PIM designs place the compute next to
the memory arrays~\cite{attacc, neupims, duplex, m2ndp}.
Such designs, however, target dense MHA or GQA, whose decode kernels run at an
OI of a few FLOPs/byte, and size their compute engines accordingly.
Retrieval-based sparse attention violates this assumption.
First, the OI is higher: MLA reconstructs every head from a single low-rank
latent per token, so one byte read feeds the arithmetic of all heads and core
attention reaches 114--228 against 16 under GQA, while DSA scores each token
over several indexer heads that share one FP8 index key, placing the indexer at
64--128 against 4 under coarse block scoring (Table~\ref{tab:models}).
Provisioned for the lower intensity, prior PNM units render the scan
compute-bound and become the bottleneck themselves.
Second, score functions, selection granularities, and indexer cadences differ
across models and change with every generation (Table~\ref{tab:op_diversity}),
which no fixed enumeration of function units can cover.

\obs{O5}\emph{Retrieval-based sparse attention still calls for near-memory
processing, but it changes what such a design must provide: the compute must
remain colocated with the KV cache and index keys, sustain a higher OI than
prior designs assume, and accommodate operations that vary across models.}

\begin{table}[t]
\centering
\caption{Design requirements for retrieval-based sparse attention at
long context.}
\setlength{\tabcolsep}{4pt}
\renewcommand{\arraystretch}{1.15}
\footnotesize
\begin{tabular}{@{}>{\raggedright\arraybackslash}p{1.75cm}|>{\raggedright\arraybackslash}p{5.9cm}@{}}
\hline
Requirement & Required characteristic \\ \hline\hline
\textbf{R1}: Capacity \newline (\obsref{O1}, \obsref{O2}, \obsref{O4}) &
Capacity per watt an order of magnitude above HBM's, keeping the batch's KV
cache and index keys resident (78--137\,GB per 1M tokens)
\\ \hline
\textbf{R2}: Bandwidth \newline (\obsref{O3}, \obsref{O5}) &
Bandwidth per watt comparable to HBM's, co-located with the index keys, so that
the resident context is scanned within the TBT SLO  
\\ \hline
\textbf{R3}: Compute \newline (\obsref{O3}, \obsref{O5}) &
Ridge point above the OI of the indexer scan to avoid a compute bottleneck 
\\ \hline
\textbf{R4}: Generality \newline (\obsref{O3}, \obsref{O5}) &
Cores programmable across all operations in Table~\ref{tab:op_diversity},
not specialized to a single mechanism, so that current and future algorithms in this domain run without hardware change
\\ \hline
\end{tabular}
\label{tab:requirements}
\end{table}

\subsection{Design Requirements}
\label{sec:bg_req}
Observations \obsref{O1}--\obsref{O5} yield four requirements, summarized in
Table~\ref{tab:requirements}.
Because datacenters provision serving systems under a fixed power budget, the
first two are stated per watt.
No existing platform exhibits all four: GPU and PIM systems satisfy R2 and R3
but not R1, host- and CXL-based expansion satisfies R1 but not R2, and SPNM
fails R4 by construction.

\section{System Design}
\label{sec:arch}

\begin{figure*}
    \centering
    \includegraphics[width=0.95\linewidth]{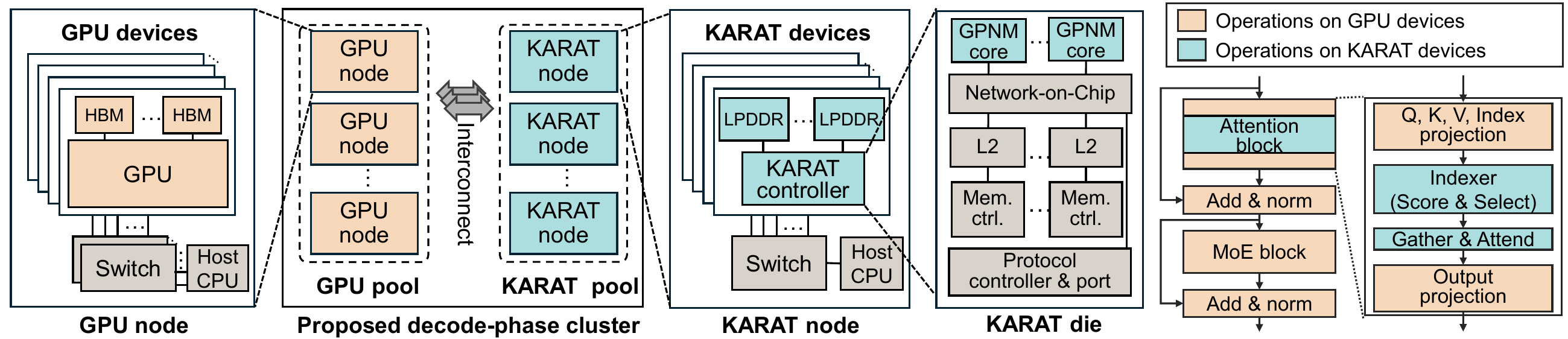}\\
    {\scriptsize\raggedright\noindent%
    \hspace*{0.14\linewidth}(a)%
    \hspace{0.16\linewidth}(b)%
    \hspace{0.17\linewidth}(c)%
    \hspace{0.115\linewidth}(d)%
    \hspace{0.16\linewidth}(e)\par}
    \caption{
        (a-d) Proposed system organization 
        (e) Execution mapping of one decode layer. Colors denote the executing
        device: projections and MoE run on GPU, while the indexer scan
        (\textsc{Score}, \textsc{Select}), \textsc{Gather}, and \textsc{Attend} run on \gpnm{}. 
    }
    \label{fig:system_overview}
\end{figure*}

\subsection{Overview}
\label{sec:arch_overview}

To meet the requirements of retrieval-based sparse attention,
we propose a heterogeneous decode-phase serving
system for long-context LLMs (Fig.~\ref{fig:system_overview}). It scales
out over a high-performance network 
(e.g., CXL~\cite{cxl_switch_isca26} and RDMA~\cite{rdma_meta})
connecting GPU nodes with \gpnm{} nodes, built from
\emph{\textbf{K}V-cache-resident \textbf{A}ccelerator for \textbf{R}etrieval-based
\textbf{AT}tention} (\gpnm{}) devices with GPNM cores.

Requests are served under prefill--decode disaggregation~\cite{splitwise, mooncake, nvidia_dynamo}.
Prefill runs on dedicated GPU nodes, and the generated KV cache is ingested into
\gpnm{} nodes over the network. Decode then partitions work by operation type:
\gpnm{} nodes hold the KV cache and index keys and execute the operations that
read them (e.g., indexer scan, \textsc{Attend}), while GPU nodes hold the model
weights and execute the compute-bound operations (e.g., QKV projection, MoE)
(Fig.~\ref{fig:system_overview}e). A \gpnm{} node comprises many (e.g., 32)
devices, each combining a high-capacity memory device (e.g., LPDDR) that stores
long-context KV caches with compute units on the \gpnm{} die that execute the
indexer scan and attention (Fig.~\ref{fig:system_overview}c,d).

Under a given power constraint, a workload-aware configuration search
(\S\ref{sec:arch_config}) determines the composition ratio between the two node
types from the workload characteristics, and KV placement
(\S\ref{sec:arch_mapper}) keeps both types utilized within the TBT SLO. Because
they execute disjoint operations within a layer, the two node types are
pipelined across micro-batches, and we propose two techniques that reduce the
resulting bubbles: \emph{opportunistic, fine-grained micro-batch scheduling}
(OFMS) hides expert all-to-all communication behind the
other micro-batch's GEMMs, and \emph{context-length-aware micro-batch
rebalancing} (CMR) balances their token counts given
the wide variance in context length.

\subsection{Execution Workflow}
\label{sec:arch_workflow}

After prefill, only the KV cache of the newly prefilled tokens is transferred to
the \gpnm{} nodes, since the context of earlier turns already resides there.
Ingestion traffic therefore scales with the number of new tokens, not the full context length.

During decode (Fig.~\ref{fig:system_overview}e), a GPU node first builds the
indexer Q and K for the new token (\textsc{Index-Build}) in a kernel fused with
the QKV projection. The indexer Q and K are sent to the \gpnm{} nodes first so
that \textsc{Score} can begin, as the rest of the attention layer depends on its
result; the QKV vectors of the new token follow in the background, overlapping
with \textsc{Score}. The \gpnm{} nodes then select the top-$k$ candidates and
gather their KV entries locally (\textsc{Select}, \textsc{Gather}), feed them to
\textsc{Attend}, and return the attention output to the GPU nodes. The GPU nodes
apply the output projection and execute the MoE layer with expert parallelism,
\emph{dispatching} tokens to the experts across GPU nodes and \emph{combining}
their outputs through all-to-all communication to form the layer output.

Offloading the attention layer thus requires communication between the two node
types, but it consumes only a small fraction of the TBT budget. Per layer and
per request, the indexer Q and K amount to $1.3$--$8.6$\,KB, the attention Q to
$16$--$49$\,KB, and the attention output to $16$--$131$\,KB, far below the
$2.4$--$4.2$\,MB that \textsc{Gather} moves within a \gpnm{} device. Summed over
all layers and charged without any overlap, these transfers take
$43$--$232$\,\textmu s per request and decoding step, below $0.25\%$ of the
100\,ms TBT SLO.

\subsection{Mapping KV Cache to \gpnm{} Devices}
\label{sec:arch_mapper}

A \gpnm{} node comprises many \gpnm{} devices, each computing only on its
local memory, so the placement of a request's KV cache determines how much
traffic crosses the switch. Splitting the $O(L)$ indexer scan across devices is
possible, but \textsc{Gather} and \textsc{Attend} touch only $O(k)$ tokens, so
fetching them remotely adds a transfer disproportionate to the short attention
itself, and concurrent gathers can congest the switch. We therefore keep each
request's context on a single device and run its indexer there as well.
Production stacks already parallelize attention this way, replicating it across
data-parallel ranks so that each rank attends over only its own requests' KV
cache while sharding the experts with expert parallelism~\cite{deepseek_v3,
sglang, vllm}. \gpnm{} applies the same request-level data parallelism at a finer granularity,
where the unit of placement is an individual \gpnm{} device.

We allocate the KV cache at page granularity, but with a locality preference:
the first page of a request goes to the least-occupied \gpnm{} device, which
becomes its \emph{home} device, and all subsequent pages follow there. Since the
per-token index keys are co-located with the KV cache, every stage from
\textsc{Score} through \textsc{Attend} therefore completes on the home device,
spilling to another device within the same node only if the home device
is exhausted---which never occurs in our evaluation, where the per-device capacity
is sufficient. 

\subsection{\gpnm{} Device Design}
\label{sec:arch_device}

Three principles follow from R1--R4 and determine the design independently of any
specific technology or process. First, because a datacenter operates under a
fixed power budget, memory technologies must be compared per watt rather than in
absolute capacity or bandwidth. Of the two, capacity/W is the more demanding,
since R1 calls for an order of magnitude beyond what HBM provides whereas R2
calls only for parity with it. Second, a device meeting R1 cannot be built as a
single large die: the memory types with high capacity/W also demand more die-edge
I/O per unit bandwidth, so capacity and bandwidth must be aggregated over many
small dies rather than provided by one. Third, the compute engine is sized to the
OI of the offloaded operations rather than to peak throughput, and must be
programmable across all five of them (R4). Any programmable core type that
satisfies the same OI requirement would serve equally well. These principles rest
on the design targets of each memory type and therefore persist across
generations.
The rest of this subsection instantiates them at today's technology node. 
Successor devices (e.g., GDDR7, LPDDR6, HBM4) preserve the same relative
characteristics and hence the same design.

\noindent\textbf{Memory device.}
Table~\ref{tab:memory_device} compares memory expansion devices of comparable
generations. DDR5 offers the highest capacity/W but only 0.23$\times$ the
bandwidth/W of LPDDR5X, too low for R2, whereas GDDR6 inverts the trade-off at
1.42$\times$ the bandwidth/W and 0.07$\times$ the capacity/W. LPDDR5X is the only
type that meets both R1 and R2, providing 19$\times$ the capacity/W of HBM3 at
0.8$\times$ its bandwidth/W. Its PHY, however, requires at least 5.9$\times$ more
shoreline per unit bandwidth than HBM~\cite{LPDDR6_shoreline, micron_hbm3e,
memexplorer}, so a fixed-size die hosts far fewer channels and a drop-in
replacement of a GPU's HBM would reduce both capacity and bandwidth. We instead
spread the memory across many smaller dies, whose perimeter shrinks more slowly
than their area and which therefore expose more aggregate shoreline for the same
total die area.

\begin{figure}
    \centering
    \parbox{0.56\linewidth}{
    \centering
        \includegraphics[width=.9\linewidth]{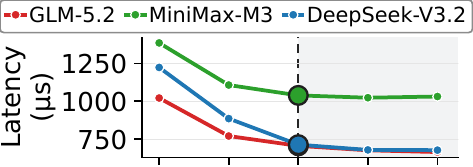}
        \\[-0.05in]
        \hspace{.455in}\scriptsize{(a)}
        \\[0.01in]
        \includegraphics[width=.9\linewidth]{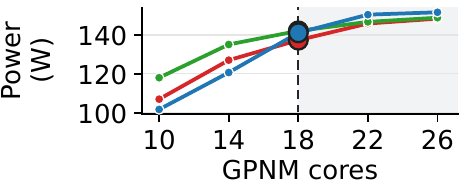}
        \\[-0.07in]
        \hspace{.455in}\scriptsize{(b)}
    }
    \parbox{0.42\linewidth}{
    \centering
        \includegraphics[width=.9\linewidth]{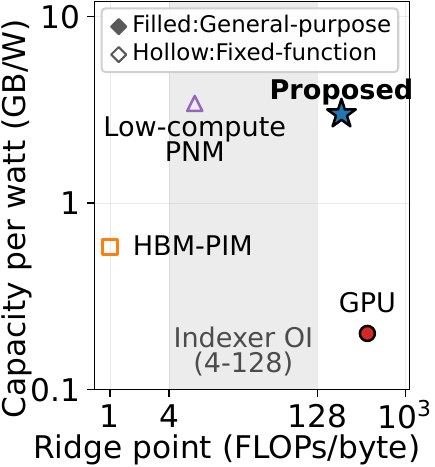}
        \\[-0.05in]
        \hspace{.23in}\scriptsize{(c)}
    }
\caption{Compute provisioning for the \gpnm{} device. Indexer scan (a) latency per layer
and (b) power as the GPNM core count varies, for the three models
of Table~\ref{tab:models} at batch size 256 with 16K context per request (4M tokens in total); the
dashed line marks the selected 18 cores. (c) Capacity per watt versus ridge point
across accelerator classes. Shading marks the indexer OI range, which a design's ridge point must exceed to
avoid a compute bottleneck (R3);\gpnm{} meets all three: sufficient capacity per watt (R1) with general-purpose
cores (R4).
\S\ref{sec:eval_power} describes the area and power models.}
    \label{fig:gpc_sweep}
\end{figure}

\noindent\textbf{Compute provisioning.}
We build the GPNM cores as a compute engine of GPU SM-like cores~\cite{nsu, tom}
and size it to the device's LPDDR5X-9600 bandwidth of 1.2\,TB/s: holding this
bandwidth fixed, we sweep the core count and measure indexer scan latency and
power (Fig.~\ref{fig:gpc_sweep}a--b) across target LLMs. 
Latency improves only up to 18 cores, beyond which additional cores add power and
area with little benefit, so we provision 18 cores for the indexer scan and
attention operations. Even with
these cores and a 15\,MB L2 cache---retained from the SM-based baseline we model
in \S\ref{sec:eval_power}---the LPDDR5X PHY and memory controllers occupy
166\,mm$^2$, or 58\% of the 286\,mm$^2$ die, confirming that the device remains
memory-centric. 
Finally, we configure the PCIe interface as $\times$4 and cap the TDP at 175\,W,
a quarter of the H200's 700\,W, so that four devices fit a single GPU's power and
lane budget.

The resulting device pairs 512\,GB of LPDDR5X at 1.2\,TB/s with a compute engine
rated at 250\,TFLOPS for FP8 and 125\,TFLOPS for BF16. At FP8, the precision in
which index keys are stored~\cite{deepseek-v3.2, glm-5.2}, its ridge point of
208\,FLOPS/byte lies above the 4--128 OI range of the indexer scan, so
the cores do not become the bottleneck (R3), and \gpnm{} is the only class in
Fig.~\ref{fig:gpc_sweep}c that meets all four requirements.

\noindent\textbf{\gpnm{} node design.}
\label{sec:arch_node}
Each \gpnm{} node comprises 32 \gpnm{} devices, a PCIe/CXL switch, and a host CPU,
provisioned to match both the accelerator TDP and the aggregate memory bandwidth
of an 8-GPU node (Table~\ref{tab:node_comparison}). Under equal power and
bandwidth, it holds 14$\times$ more memory at roughly half the BF16
throughput---a trade that favors the offloaded operations, which are memory bandwidth- and capacity-bound.
Each node connects to the GPU pool through scalable interconnect,
and its host CPU manages task scheduling and execution within the node.

\begin{table}
      \centering
      \caption{Comparison of memory expansion devices across memory technologies,
  with per-watt metrics derived from the specifications
  in~\cite{cxl-pnm}. Values are normalized to the LPDDR5X-based device.}
      \scriptsize
    \begin{threeparttable}
      \begin{tabular}{c||cccc}
      \hline
    & DDR5 & GDDR6 & HBM3 & LPDDR5X$^{*}$ \\ \hline \hline
  Capacity (GB) & 512 & 32 & 80 & 512 \\  \hline
  BW (GB/s) & 89.6 & 1536 & 4198.4 & 1126 \\ \hline
  Normalized power & 0.35 & 0.96 & 3.00 & 1.00 \\ \hline
  Normalized capacity/W & 2.86 & 0.07 & 0.05 & 1.00\\ \hline
  Normalized BW/W & 0.23 & 1.42 & 1.24 & 1.00 \\ \hline
      \end{tabular}
    \begin{tablenotes}[flushleft]\scriptsize
   \item[] $^{*}$LPDDR5X-8533; the \gpnm{} device uses LPDDR5X-9600 at 1.2\,TB/s.
    \end{tablenotes}
    \end{threeparttable}
      \label{tab:memory_device}
\end{table}

\begin{table}
    \centering
    \caption{Comparison of example GPU and \gpnm{} nodes. The host CPU is identical for both node type and omitted for brevity.}
    \scriptsize
    \begin{tabular}{c||cc}
    \hline
     & GPU node & \gpnm{} node \\ \hline \hline
     Devices & 8 GPUs & 32 \gpnm{} devices\\ \hline
     Power (kW) & 5.6 (8$\times$0.7) & 5.6 (32$\times$0.175) \\ \hline
     Memory capacity (TB) & 1.13 (8$\times$0.141) & 16 (32$\times$0.5) \\ \hline
     Memory bandwidth (TB/s) & 38.4 (8$\times$4.8) & 38.4 (32$\times$1.2) \\ \hline
     BF16 TFLOPS & 7912 (8$\times$989) & 4000 (32$\times$125) \\ \hline
    \end{tabular}
    \label{tab:node_comparison}
\end{table}

\subsection{Opportunistic, Fine-grained Micro-batch Scheduling}
\label{sec:arch_sched}

While micro-batching and pipelining overlap GPU execution with \gpnm{} execution, significant pipeline bubbles can still arise on the GPUs because of the communication phases interleaved between GPU kernels within a micro-batch.
Prior micro-batching approaches for offloading the attention layer~\cite{neupims, duplex, megascale-infer} partition the operations of an LLM into two sets of successive kernels, one compute-bound and the other memory-bound. The memory-bound kernels run on the memory-side compute units, while the GPU executes the compute-bound sequence—output projection, FFN/MoE, and QKV projection. We refer to this approach as \emph{static, coarse-grained micro-batch scheduling} (SCMS).

\begin{figure}
    \centering
    \includegraphics[width=0.99\linewidth]{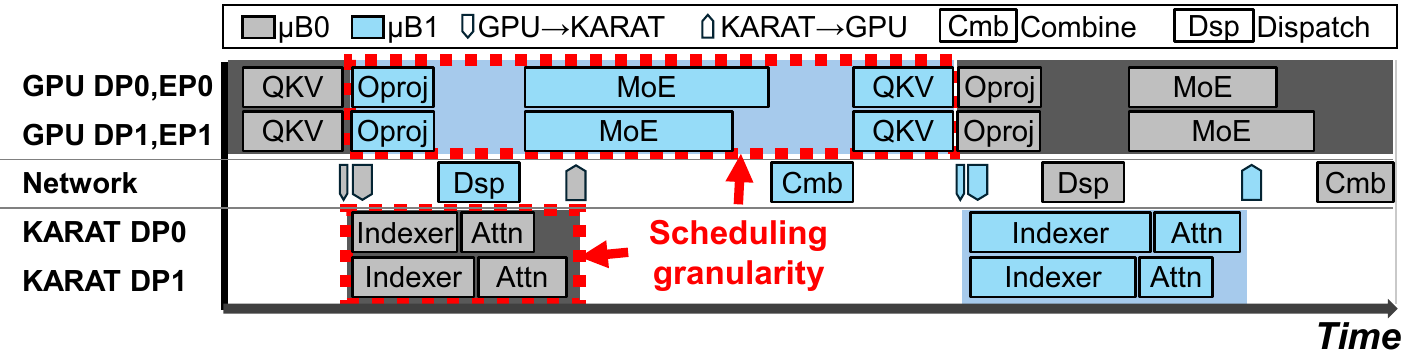}
    \vspace{-.30in}
    \\
    \centerline{\scriptsize{(a) Baseline SCMS, GPU-bound scenario}}
    \vspace{0.05in}
    \includegraphics[width=0.99\linewidth]{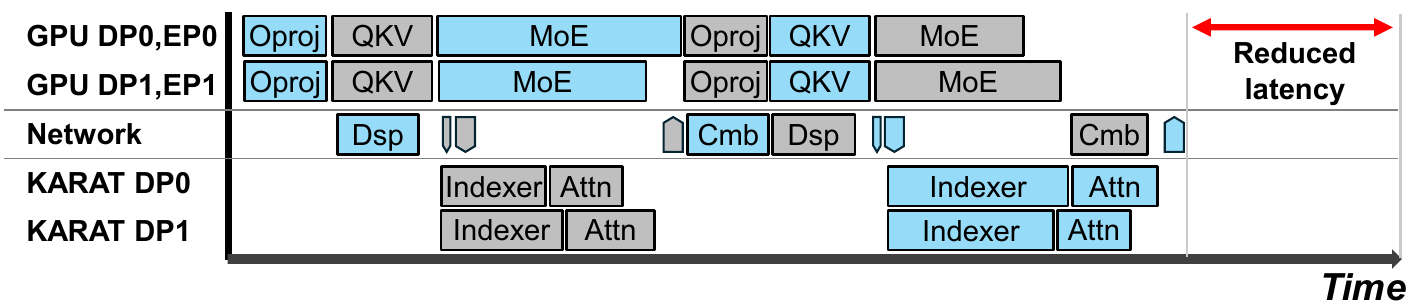}
    \vspace{-.30in}
    \\
    \centerline{\scriptsize (b) Proposed OFMS, GPU-bound scenario}
    \vspace{0.05in}
    \includegraphics[width=0.99\linewidth]{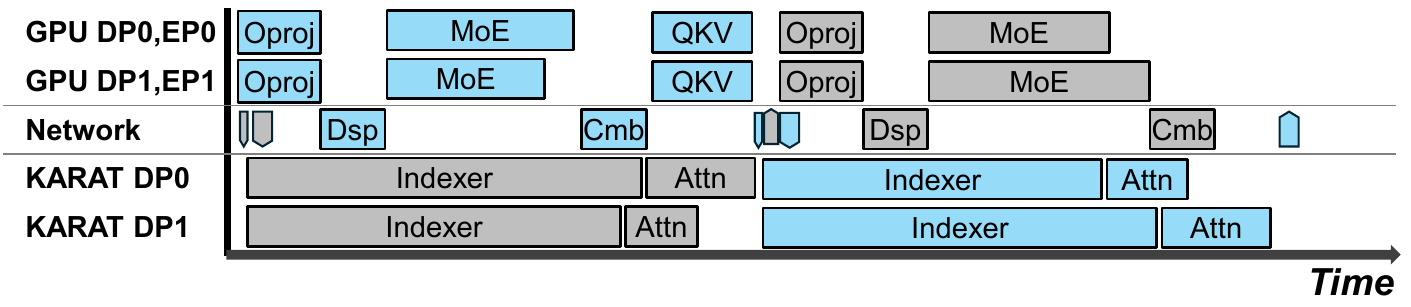}
    \vspace{-.30in}
    \\
    \centerline{\scriptsize (c) Proposed OFMS, \gpnm{}-bound scenario}
    \caption{Timeline showing different micro-batch scheduling approaches with two micro-batches (\ub{0}, \ub{1}).}
    \label{fig:gantt_schedule}
\end{figure}

A key limitation of SCMS is that, when expert parallelism spans multiple nodes, the communication for dispatching and combining leaves the GPUs idle for a substantial fraction of the time (Fig.~\ref{fig:gantt_schedule}a). Moreover, this communication latency grows with the batch size, which itself scales with total system memory capacity (\S\ref{sec:bg_dpep}). The idle time is especially wasteful when the \gpnm{} device finishes \textsc{Attend} early—either because severe expert skew prolongs the MoE stage on the GPU, or because the current attention layer has no indexer as a result of cross-layer index reuse~\cite{glm-5.2}.

To overcome this limitation, we propose \emph{opportunistic, fine-grained micro-batch scheduling} (OFMS), illustrated in Fig.~\ref{fig:gantt_schedule}b. As soon as \textsc{Attend} completes for a micro-batch (\ub{0}) and its output reaches the GPUs, we launch its output projection, provided the GPUs are idle waiting on the MoE combining of the other micro-batch (\ub{1}). Its MoE dispatch then follows immediately, overlapping with the QKV projection that follows the earlier combining (\ub{1}). GPU idle time is thereby substantially reduced.

Opportunistic overlapping is not always possible (Fig.~\ref{fig:gantt_schedule}c): under a long context, the indexer scan and \textsc{Attend} may be slow enough that the output of \ub{0} does not arrive before the QKV projection of \ub{1} begins. OFMS therefore reorders only when the opportunity actually exists and otherwise falls back to the SCMS order, adapting to the runtime latency of each operation rather than committing to a fixed pattern in advance. It thus does not perform worse than SCMS, and our evaluation shows that this flexibility improves performance across a range of LLMs and workloads.

OFMS can be implemented in a fully decentralized manner. Each data-parallel rank schedules its own micro-batches, and within a rank each worker (a GPU or a \gpnm{}) dispatches operations from its own queue, switching micro-batches only when the current operation blocks on a transfer. No global scheduler or shared state is needed, since each decision inspects only the local queue head. Nor does the schedule add any cross-rank synchronization beyond the all-to-all dispatch and combine that expert parallelism already performs.

\subsection{Context-length-aware Micro-batch Rebalancing} 
\label{sec:cmr}

Each batch is split into two micro-batches that pipeline the GPU and the
\gpnm{} devices. On the \gpnm{} devices, the execution time of a micro-batch
scales with its aggregate context length, so an imbalance in aggregate context
length between the two micro-batches lengthens the interval during which the
GPU waits for the longer one.
Prior approaches~\cite{neupims,
megascale-infer, exegpt} balance the micro-batches by request count and
overlook the aggregate context length; moreover, they do not revise the
partition, since newly admitted requests are appended to the existing
composition. The wide variance in context length across requests can therefore leave the two micro-batches unbalanced.

We propose \emph{context-length-aware micro-batch rebalancing} (CMR), which
repartitions the requests across the two micro-batches whenever a request
finishes or is admitted. Let $R_d$ denote the requests whose KV cache resides on
\gpnm{} device $d$, let $L(r)$ denote the context length of request $r$, and let
$\ell(m,d)=\sum_{r\in m\cap R_d} L(r)$ denote the aggregate context length that
micro-batch $m$ places on $d$. Algorithm~\ref{alg:cmr} empties every micro-batch
(line~\ref{ln:cmr-reclaim}) and partitions each $R_d$ from scratch in order
of decreasing context length (line~\ref{ln:cmr-sort}), assigning each request to
the micro-batch of smallest $\ell(m,d)$ (line~\ref{ln:cmr-argmin}); ties are
resolved by micro-batch size $|m|$. CMR reassigns every request rather than only
the newly admitted one, which corrects imbalance that has already accumulated.

CMR incurs little overhead. It moves no data between \gpnm{} devices, since it
changes only the micro-batch a request belongs to and not the device that holds
its KV cache. Its computation is one sort of $R_d$ per device on each rebalance.
Between rebalances, $\ell(m,d)$ drifts by at most the difference in request
count on $d$, negligible relative to contexts of hundreds of thousands of
tokens, so no further trigger is needed.

\begin{algorithm}[t]
\small
\caption{CMR. $M$: the two micro-batches; $D$: \gpnm{} devices;
       $R_d$: requests whose KV resides on $d$; $L(r)$: context length of $r$;
       $|m|$: number of requests in $m$;
       $\ell(m,d)=\sum_{r\in m\cap R_d}L(r)$.}
\label{alg:cmr}
\begin{algorithmic}[1]
\State $m \gets \emptyset \;\; \forall m \in M$
     \label{ln:cmr-reclaim} \Comment{empty every micro-batch}
\ForAll{$d \in D$}
\ForAll{$r \in \textsc{SortDesc}(R_d,\, L)$} \label{ln:cmr-sort}
       \Comment{decreasing ctx. len.}
  \State $m^{*} \gets \arg\min_{m \in M}
          \big(\ell(m,d),\; |m|\big)$ \label{ln:cmr-argmin}
  \State $m^{*} \gets m^{*} \cup \{r\}$
\EndFor
\EndFor
\end{algorithmic}
\end{algorithm}

\subsection{Workload-Aware System Configuration Search}
\label{sec:arch_config}

A \ndp{} system is heterogeneous: GPU and \gpnm{} nodes have distinct
characteristics, so combining them in the right proportion is essential.
The best proportion also depends on the model and on the degrees of parallelism
it admits, which affect performance substantially.
To explore this design space, we search for the system configuration that
maximizes throughput under iso-power and SLO constraints, using a workload
profile and a measured operation-latency lookup table (LUT) for the model.
The search space consists of
    (i) the ratio of GPU to \gpnm{} nodes,
    (ii) the expert parallelism degree, and
    (iii) the attention tensor parallelism degree.
For each configuration, we compute the throughput at the largest batch size that
meets the SLO constraint. The space is discrete and contains only hundreds of
points, so an exhaustive grid search completes in minutes.

\section{Evaluation}
\label{sec:eval}

\subsection{Methodology}
\label{sec:method}

\noindent\textbf{System configuration.}
We evaluated the decode pool under prefill--decode disaggregation, where the
two pools were provisioned independently, as in deployment
practice~\cite{splitwise, distserve, nvidia_dynamo, mooncake}. All workloads
ran at 8-node scale; node and device configurations were as given in
Tables~\ref{tab:configurations} and~\ref{tab:node_comparison}. The baseline
used 64 H200 GPUs (8 nodes $\times$ 8 GPUs), and for every other system under
comparison we chose, among all valid node-type combinations, the
configuration whose aggregate TDP was closest to that of the baseline. Both
node types provided 400\,GB/s of aggregate per-node inter-node bandwidth,
matching DGX H200~\cite{dgx_h200}. For each system, we then selected the
expert-parallel (EP) and tensor-parallel (TP) degrees per model and workload
profile using the search method in \S\ref{sec:arch_config}, following
industry practice~\cite{aiconfigurator, nvidia_dynamo_sla_planner,
nvidia_dynamo_planner_design}.

\noindent\textbf{Baselines.}
The baselines cover the alternatives examined in \S\ref{sec:bg_fail}.
\texttt{GPU-only} serves all layers on H200 GPUs, so the KV cache and index
keys must fit in HBM. \texttt{HiSparse}~\cite{hisparse} adds capacity by
offloading the KV cache to host DRAM (2\,TB DDR5 per GPU node) and caching the
selected entries in HBM; we assumed its reported maximum hit rate of 95\% and
charged transfers at data size over PCIe bandwidth, both of which favor it.
\texttt{Duplex}~\cite{duplex} and \texttt{AttAcc}~\cite{attacc} place
special-purpose compute in memory, the former by adding PIM units to the GPU
and the latter as a PIM-based heterogeneous system. Neither supports MLA, so
\texttt{AttAcc} appears only for MiniMax-M3, and \texttt{Duplex} accelerates
only the MoE layers on the two MLA models.
\texttt{SPNM}~\cite{cxl-pnm, cxl-llm-pact25} places special-purpose compute in
LPDDR5X-based CXL memory; we enabled its micro-batching wherever it improved
throughput. Table~\ref{tab:configurations} lists the parameters we used.
For training-free sparse attention, we evaluated one method from each selection
family in Table~\ref{tab:op_diversity}: page-based Quest~\cite{quest},
clustering-based STARC~\cite{starc}, and hash-based MagicPIG~\cite{magicpig}.
MagicPIG targets the CPU, but the GPU outperformed it at this scale, so we used
the GPU. 
When an evaluated device could not execute an operation required by a
workload, we ran that operation on the GPU.

\noindent{\bf Models.}
We evaluated three state-of-the-art models with model-native sparse attention,
one from each mechanism family classified in \S\ref{sec:bg_sparse}:
GLM-5.2~\cite{glm-5.2}, MiniMax-M3~\cite{minimax-m3}, and
DeepSeek-V3.2~\cite{deepseek-v3.2} (Table~\ref{tab:models}).
Weights are FP8 for DeepSeek-V3.2 and BF16 for the other two, and the KV cache
is BF16 throughout. Index keys follow each indexer's definition: FP8 for the
two DSA models and BF16 for MiniMax-M3. For training-free sparse attention, we used LLaMA3-70B~\cite{llama3} in BF16 as the base model.

\noindent{\bf Workloads.}
We used three real agentic traces (Table~\ref{tab:workloads}):
SWE-Bench-Pro (SWE-Pro)~\cite{swepro}, 
SWE-chat~\cite{swe-chat}, and KV-cache-tester (KVCT)~\cite{kvct}.  
Each run is closed-loop at a fixed concurrency $N$: $N$
sessions are in flight at all times, a session issues its next turn only after
the previous one completes, and a session that ends is immediately replaced by a
new one drawn from the trace with replacement, so the workload follows the trace
distribution regardless of run length. The initial $N$ sessions start at a random
turn to avoid a cold-start transient, while replacements start at their first
turn. We swept $N$ and report the highest decode throughput (tokens/s) after warm-up 
that meets a 100\,ms P99 TBT SLO, normalized by TDP as system power cannot be equalized.

\begin{table}
    \centering
    \begin{threeparttable}
    \caption{Evaluated device configuration.}
    \label{tab:configurations}
    \footnotesize
    \begin{tabular}{c||c}
    \hline
         Component & Configuration \\  \hline \hline
        \multirow{2}{*}{H200 GPU~\cite{h200}}
            & 132 SMs @ 1.98 GHz, 141 GB HBM3e @ 4.8 TB/s, \\
            &  PCIe5/CXL $\times$16 (64 GB/s), TDP: 700 W\\ \hline
       \multirow{2}{6em}{Duplex~\cite{duplex}}
            & H200 + Logic-PIMs ($\times$4 internal BW, 153 TFLOPS),\\
            & TDP: 1000W\\ \hline
        \multirow{2}{6em}{AttAcc~\cite{attacc}}
            & 80-channel HBM3-PIMs with 3.35TB/s, 67.2GB$^{*}$\\
            & PCIe5/CXL $\times$4 (16 GB/s), TDP: 120W\\ \hline
        \multirow{3}{6em}{SPNM~\cite{cxl-pnm, cxl-llm-pact25}}
            & 8~TFLOPS (BF16) @ 1~GHz\\
            & LPDDR5X (1.2TB/s, 512GB from 64ch)\\
            & PCIe5/CXL $\times$4 (16 GB/s), TDP: 150W\\ \hline
        \multirow{4}{6em}{\gpnm{}}
            & 18 GPNM cores @ 1.83 GHz, \\
            & 250 / 125 TFLOPS (FP8 / BF16), 15~MB L2 cache \\
            & LPDDR5X (1.2TB/s, 512GB from 64ch)\\
            & PCIe5/CXL $\times$4 (16 GB/s), TDP: 175W\\ \hline
    \end{tabular}
  \begin{tablenotes}[flushleft]\scriptsize
  \setlength{\itemsep}{0pt}\setlength{\parskip}{0pt}
 \item[] $^{*}$The AttAcc work~\cite{attacc}
did not address the capacity impact of PIM, but we assume the HBM DRAM die
remains the same to preserve yield while bank area is reduced to accommodate PIM
units with a 14.59\% area overhead.
Assuming the banks occupy 90\% of the die (as TSVs use
$\sim$10\%~\cite{hbm3_isscc22}), the memory capacity is reduced by 16\%
from the original 80\,GB HBM3.
  \end{tablenotes}
    \end{threeparttable}
\end{table}

\begin{table}[t]
  \centering
  \begin{threeparttable}
  \caption{Workload characteristics.}
  \label{tab:workloads}
  \footnotesize
  \setlength{\tabcolsep}{5pt}
  \begin{tabular}{l||r|r|l|l|l}
  \hline
  Workload & Avg. turns & Avg. ctx. & Peak ctx. & Pref./Dec. & Reuse \\ \hline \hline
  SWE-pro   & 30 & 63K  & 295K  & 324/263 & 97\% \\ \hline
  SWE-chat  & 39 & 81K  & 1.06M & 387/47  & 97\% \\ \hline
  KVCT      & 48 & 144K & 894K  & 647/218 & 94\% \\ \hline
  \end{tabular}
  \begin{tablenotes}[flushleft]\scriptsize
  \setlength{\itemsep}{0pt}\setlength{\parskip}{0pt}
  \item[] Real agentic traces: SWE-pro~\cite{swepro} is a Codex
    trace~\cite{swepro_codex}; SWE-chat~\cite{swe-chat} and KVCT~\cite{kvct}
    from Claude Opus. Ctx.\ is the context length in tokens, Pref./Dec.\ the
    median prefill and decode tokens per turn, and Reuse the fraction of KV
    cache read from earlier turns.
  \end{tablenotes}
  \end{threeparttable}
\end{table}

\begin{figure*}
    \centering
    \includegraphics[width=.95\linewidth]{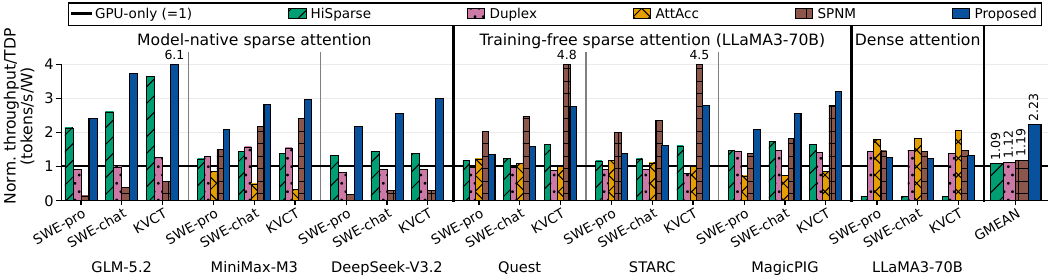}
  \caption{Throughput per TDP under a 100\,ms P99 TBT SLO, normalized to
  \texttt{GPU-only} for the same model and trace. Model-native sparse attention
  uses GLM-5.2, MiniMax-M3, and DeepSeek-V3.2; training-free and dense attention
  use LLaMA3-70B. Bars above the axis limit are clipped and labeled.
  \texttt{AttAcc} does not support MLA, so it appears only for MiniMax-M3 among
  the model-native models and is excluded from the geometric mean.}
    \label{fig:eval_throughput}
\end{figure*}

\noindent\textbf{Simulator.}
We used an in-house event-driven simulator for LLM serving, built following the
approach of LLMServingSim~\cite{llmservingsim}. The simulator obtained the
latency of each operation from a lookup table populated by offline profiling.
We profiled H200 operation latencies on real hardware. For \gpnm{} we used an
H200 MIG \texttt{1g.35gb} instance~\cite{nvidia_mig_profiles}, which provides
the same memory bandwidth, and scaled the SM count~\cite{nvidia_mps}. 
We modeled communication latency with
ASTRA-sim~\cite{astra-sim2}, assuming CXL-switched inter-node links with a
600\,ns round-trip latency~\cite{pond, octopus}. For MoE layers, we assigned
tokens to experts by uniform sampling; \S\ref{sec:eval_sensitivity} examines the
effect of skew. For prior work, we used the simulators released by the
respective authors: \texttt{AttAcc}~\cite{attacc_simulator}, \texttt{Duplex} and
\texttt{SPNM}~\cite{llmsimulator}, and \texttt{STARC}~\cite{starc_artifact}.
Each run spanned 250K generated tokens or 300\,s of simulated time, whichever
occurred first, and we excluded the first 20\% as warmup.

\subsection{Performance Results}
\label{sec:eval_main}

\noindent\textbf{Model-native sparse attention.}
\ndp{} reached the highest throughput per TDP on every model and trace,
$2.09$--$6.13\times$ that of \texttt{GPU-only} (Fig.~\ref{fig:eval_throughput}).
The gain came from larger batches: offloading the KV cache and index keys to
\gpnm{} devices freed GPU HBM, so more sessions ran concurrently and the MoE and
dense FC layers operated at a larger batch
(Fig.~\ref{fig:eval_latency_breakdown}). The speedup grew with context length,
from $2.09$--$2.42\times$ on SWE-pro to $2.98$--$6.13\times$ on KVCT, whose
sessions are the longest.
\ndp{} outperformed \texttt{HiSparse} by $1.14$--$2.18\times$. The gap was
narrowest on GLM-5.2, whose cross-layer index reuse fixes the selection for four
layers at a time, letting the host-side gather overlap with compute; on the
other two models the gather follows the indexer in every layer and cannot be
hidden.
\ndp{} also outperformed \texttt{Duplex} by $1.61$--$4.83\times$.
\texttt{Duplex} adds PIM compute to the GPU but no capacity, so its concurrency
stayed at the \texttt{GPU-only} level, and its PIM units are provisioned for an
OI of 8, below the core attention of all three models ($16$ to $228$).
\texttt{AttAcc} supports only MiniMax-M3, where \ndp{} led by
$2.46$--$9.03\times$. Its 2.2\,TB capacity per node is well below the 16\,TB of
a \gpnm{} node, so concurrency fell as context grew, and its PIM compute is
likewise provisioned for low OI.
\texttt{SPNM} has the same LPDDR5X capacity as \gpnm{}, but fell below
\texttt{GPU-only} on GLM-5.2 and DeepSeek-V3.2 ($0.14$--$0.57\times$). Its
8\,TFLOPS slowed down the indexer scan of OI 64 or 128. On
MiniMax-M3, whose indexer runs at OI 4, it reached $1.50$--$2.42\times$, and
\ndp{} led by $1.23$--$1.39\times$. 
The \texttt{SPNM} results show that increasing capacity is not sufficient on its
own. A device must satisfy both the capacity and the compute throughput the
workload demands.

\noindent\textbf{Training-free sparse attention.}
\ndp{} supported all three \textsc{Select} variations and improved over
\texttt{GPU-only} by $1.36$--$3.21\times$. The other platforms fell back to the
GPU for operations they cannot perform, which left their PIM and PNM units
underutilized. The \textsc{Score} operations of Quest and STARC are dot products
followed by ranking, which \texttt{SPNM} supports, and only the periodic
re-clustering of STARC falls back to the GPU. \texttt{SPNM} therefore achieved
$2.00$--$4.80\times$ speedups on these two methods, above \ndp{}. MagicPIG
instead hashes each query with LSH, looks up the matching buckets, and samples
from them. \texttt{SPNM} can perform none of these, so all of its selection ran
on the GPU, and \gpnm{} achieved the highest improvement of $2.09$--$3.21\times$.

\noindent\textbf{Dense attention.}
For dense attention, \ndp{} achieved a $1.24$--$1.32\times$ speedup.
\texttt{AttAcc} was highest at $1.79$--$2.05\times$, as dense attention is the
low-OI GEMV it was built for, and \texttt{Duplex} and \texttt{SPNM} achieved
$1.38$--$1.46\times$ and $1.43$--$1.47\times$, respectively. \texttt{HiSparse}
instead slowed down to $0.12$--$0.13\times$, because without sparsity every
token crosses the host link.

\noindent\textbf{Summary.}
Across all workloads and traces, \gpnm{} achieved a geometric mean speedup of
$2.23\times$ in throughput per TDP over \texttt{GPU-only}. \texttt{SPNM},
\texttt{Duplex}, and \texttt{HiSparse} achieved $1.19\times$, $1.12\times$, and
$1.09\times$, respectively. \texttt{SPNM} has the same LPDDR5X capacity as
\gpnm{}, so \gpnm{}'s $1.87\times$ advantage comes from the compute placed next
to that capacity.

\subsection{Iteration Time Breakdown}
\label{sec:eval_breakdown}
\begin{figure}
\centering
    \includegraphics[width=.99\linewidth]{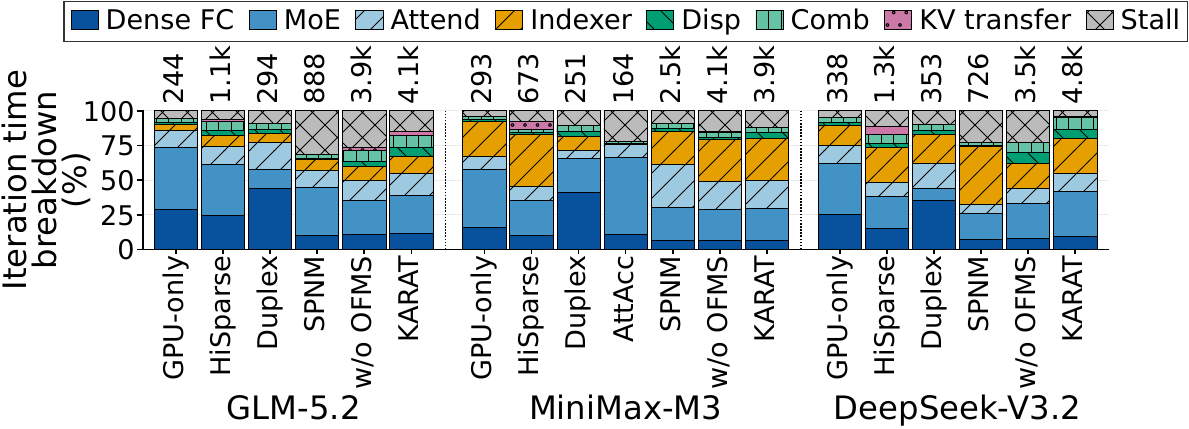}
    \vspace{-.15in}
    \caption{Decode iteration time breakdown on the KVCT trace at the
  highest-throughput operating point of each configuration under the 100\,ms P99
  TBT SLO. Numbers above the bars give the decode concurrency ($k=10^3$).}
    \label{fig:eval_latency_breakdown}
\end{figure}

Fig.~\ref{fig:eval_latency_breakdown} breaks down decode time on the KVCT trace
at the highest-throughput operating point under the SLO, with the sustained
concurrency above each bar. \texttt{GPU-only} sustained $253$--$339$ sessions,
and the dense FC and MoE layers accounted for $58$--$74\%$ of the iteration, as
a small batch cannot amortize their weight reads. \gpnm{} sustained
$4{,}017$--$5{,}055$ sessions, and those layers fell to $29$--$42\%$. The gain
therefore came from serving more sessions at once.

\texttt{SPNM} sustained $733$--$2{,}539$ sessions despite having the same
LPDDR5X capacity as \gpnm{}. The indexer and stall time together accounted for
$33$--$65\%$ of its iteration, and the indexer alone for $42\%$ on
DeepSeek-V3.2, whose indexer runs at an OI of 128. The resulting iteration time
leaves no room for a larger batch under the TBT SLO, so its capacity cannot be
translated into concurrency.

\texttt{Duplex} offloaded the MoE layers to its PIM units, but its concurrency
remained at $254$--$354$ sessions, as it adds no capacity. \texttt{AttAcc}
sustained only $164$ sessions on MiniMax-M3, fewer than \texttt{GPU-only}.
\texttt{HiSparse} sustained $674$--$1{,}340$ sessions, held below \gpnm{} by
host memory capacity.

\subsection{Ablation and Sensitivity Analysis}
\label{sec:eval_sensitivity}

\begin{figure}
    \centering
    \includegraphics[width=0.9\linewidth]{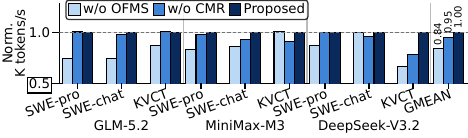}
    \caption{Performance impact of disabling OFMS and CMR.}
    \label{fig:ablation}
\end{figure}

\noindent\textbf{Scheduling.}
Disabling OFMS, which reverts the schedule to the static, coarse-grained order
of \S\ref{sec:arch_sched}, reduced throughput by $16\%$ in geometric mean
(Fig.~\ref{fig:ablation}). It also increased stall time
(Fig.~\ref{fig:eval_latency_breakdown}), and the throughput loss was largest
where that increase was largest. On DeepSeek-V3.2 with KVCT, stall time grew
from $4\%$ of the iteration to $23\%$ and throughput fell to $0.67\times$.
Disabling CMR reduced throughput by $5\%$ in geometric mean, most pronounced on
DeepSeek-V3.2 with KVCT at $21\%$, and the effect scales with the variance of
context length within a micro-batch. Both mechanisms raised the
utilization of the GPU and the \gpnm{} devices together.
Disabling OFMS reduced GPU utilization by $8.5$--$10.4\%$ and \gpnm{}
utilization by $9.4$--$13.6\%$, and disabling CMR reduced them by at most
$7.2\%$ and $6.3\%$, respectively.

\begin{figure}
\centering
\includegraphics[width=0.90\linewidth]{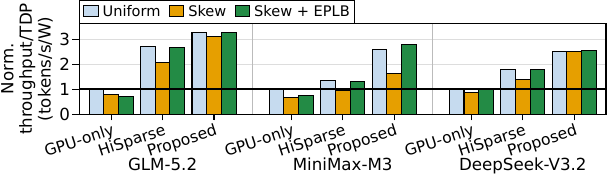}
    \caption{Normalized throughput per TDP at each configuration's best-throughput
  operating point. Each bar is the geometric mean over three agentic traces.}
    \label{fig:MoESkew}
\end{figure}

\noindent\textbf{Expert load skew.}
So far we assumed uniform expert routing. We now evaluated \ndp{} under the
routing distribution measured on the coding workload of a public trace
dataset~\cite{patterns-chaos}, the most skewed workload in that dataset
(Fig.~\ref{fig:MoESkew}). Skewed routing lowered throughput for every
configuration, and \ndp{} still outperformed \texttt{GPU-only} by
$2.4$--$3.9\times$ and \texttt{HiSparse} by $1.2$--$1.8\times$. The Expert
Parallelism Load Balancer (EPLB)~\cite{deepseek-v3.2} replicates the popular
experts and places one redundant expert per EP rank. It restored \ndp{} and
\texttt{HiSparse} to their throughput under uniform routing on all three models.
\texttt{GPU-only} recovered only on DeepSeek-V3.2 and reached $0.75\times$ on
MiniMax-M3. On GLM-5.2 it dropped further, from $0.80\times$ under skewed
routing to $0.72\times$, because the redundant experts consume the HBM capacity
that \texttt{GPU-only} lacks.

\begin{figure}
    \centering
    \includegraphics[width=0.95\linewidth]{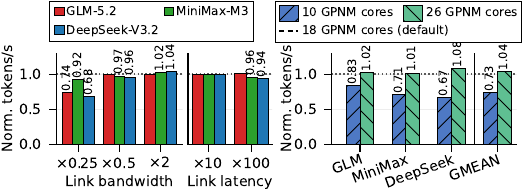}
    \\[-0.05in]
    {
    \scriptsize\raggedright\noindent%
    \hspace*{0.24\linewidth}(a)%
    \hspace{0.16\linewidth}(b)%
    \hspace{0.3\linewidth}(c)
    \par
    }
    \caption{Sensitivity of \ndp{} to (a) GPU--\gpnm{} link bandwidth, (b)
  GPU--\gpnm{} link latency, and (c) \gpnm{} core count, across three models and
  the geometric mean of the three agentic traces. Throughput is normalized to the
  default configuration.}
    \label{fig:net_sm_sens}
\end{figure}

\noindent\textbf{Link and core count.}
We swept the GPU--\gpnm{} link and the \gpnm{} core count around the default
configuration (Fig.~\ref{fig:net_sm_sens}). For the link, we varied only the
bandwidth and latency of the transfers between the GPU nodes and the \gpnm{}
nodes, leaving the expert all-to-all at its default setting, so that the sweep
only affects the path \ndp{} introduces. 
Halving the
bandwidth cost at most $4\%$ of the throughput, and doubling it resulted in a
gain of at most $4\%$. Quartering it cost $8$--$32\%$. Raising the latency by
$10\times$ had a negligible effect, and raising it by $100\times$ cost at most
$6\%$ of the throughput. Adding 8 cores to the default of 18 resulted in a
$3.8\%$ throughput gain in geometric mean, and DeepSeek-V3.2 gained the most, at
$8\%$, as it runs at the highest OI (Table~\ref{tab:models}). Reducing the core
count by 8 cost $27\%$ of the throughput in geometric mean, as 10 cores cannot
sustain the retrieval indexer at its OI. The resources within the device are therefore provisioned to match the intensity
of the operations it runs.

\begin{figure}
    \centering
    \includegraphics[width=.95\linewidth]{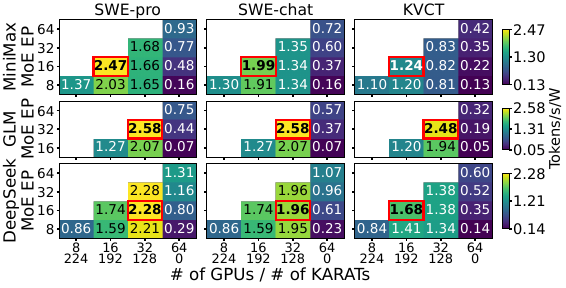}
\caption{Iso-power (44.8\,kW) design space exploration over the GPU-to-\gpnm{}
device split (x-axis) and the MoE EP degree (y-axis). Cells report decode
tokens/s/W under a 100\,ms TBT SLO on a per-panel color scale, with the red box
marking the optimum. The rightmost column (64/0) is the all-GPU baseline, and
all other cells offload the indexer scan and attention to \gpnm{}. GLM-5.2 omits
EP\,8, as its weights exceed HBM.}
\label{fig:eval_dse}
\end{figure}

\subsection{System Design Space Exploration}
\label{sec:eval_dse}

We searched a grid of four GPU-to-\gpnm{} device count splits and four MoE EP
degrees for each model and workload under a 44.8\,kW iso-power budget and the
100\,ms TBT SLO, following \S\ref{sec:arch_config} (Fig.~\ref{fig:eval_dse}).
The all-GPU column reached its highest throughput per TDP at EP\,64 on every
model and workload, as distributing the expert weights across more GPUs is the
only way to free HBM for the KV cache. Once \gpnm{} devices hold the KV cache,
the best-performing configuration moved to EP\,16--32, where the expert
all-to-all carries less data, and delivered $1.7$--$7.8\times$ the throughput
per TDP of the all-GPU column. On DeepSeek-V3.2 the best configuration also
shifted toward more \gpnm{} devices as the average context length grew, from 32
GPUs with 128 \gpnm{} devices on SWE-pro and SWE-chat, whose average contexts
are 63K and 81K, to 16 GPUs with 192 \gpnm{} devices on KVCT at 144K. The other
two models kept the same split across all three workloads.

\subsection{\ndp{} Device Area and Power Estimation}
\label{sec:eval_power}

\noindent\textbf{Area.}
We estimated the area of the \gpnm{} device from the GH100 die
breakdown~\cite{h100_die_shot}, assuming the same 4\,nm process. The 18
\gpnm{} cores (SMs) occupy 86.2\,mm$^2$ at 4.79\,mm$^2$ each, the 15\,MB L2
cache 31.2\,mm$^2$, and the PCIe PHY 2.85\,mm$^2$. We estimated the LPDDR5X PHY
and memory controller at 166\,mm$^2$ from a die shot~\cite{m2_die_shot}. The
device totals 286\,mm$^2$, $35\%$ of the 814\,mm$^2$ GH100 die, of which the
LPDDR5X PHY and memory controller account for $58\%$.

\noindent\textbf{Power.}
We measured the dynamic power of the attention and retrieval indexer kernels on
an H200 MIG \texttt{1g.35gb} instance with no other instance active, reading the
power with \texttt{nvidia-smi} and taking the difference between load and idle.
We then replaced the HBM3e energy with LPDDR5X energy using the measured memory
traffic, the HBM energy breakdown~\cite{fine_hbm}, the LPDDR5X JEDEC
specification~\cite{jedec_lpddr5}, and product data~\cite{micron_lpddr5x}. The
LPDDR5X PHY adds ${\sim}63\%$ to the energy per accessed
bit~\cite{hotchips_samsung}. For static power, we scaled the measured H200 idle
power by the die area ratio without subtracting any H200-specific block,
yielding an upper bound. The maximum power we measured was 147.7\,W, and we
conservatively assume 175\,W TDP, a quarter of the H200 TDP.

\section{Related Work}
\label{sec:related}

\subsection{LLM Serving Systems}
 Prefill--decode disaggregation provisions the two phases on separate
pools~\cite{splitwise, distserve}, and DynamoLLM~\cite{dynamo_llm} further
provisions them by request length class using an output-length
predictor~\cite{output_length_predictor}; \ndp{} targets the decode pool, where
the capacity limit binds. For MoE models, expert parallelism shards the experts
across devices: DeepEP~\cite{deepep} optimizes the all-to-all dispatch and
combine, MegaScale-Infer~\cite{megascale-infer} disaggregates the attention and
MoE layers into separate pools, and serving frameworks run attention
data-parallel while the experts stay expert-parallel~\cite{sglang, vllm,
tensorrt_llm}. These systems partition work across homogeneous GPU pools, so
every pool carries the same balance of memory and compute, and the KV cache and
index keys compete with the model weights for HBM capacity.

\subsection{Sparse Attention Algorithms}
Sparse attention reduces the work of a decoding step by attending to a subset of
the context. Eviction-based methods bound the footprint by discarding KV
pairs~\cite{h2o, snapkv, streaming-llm}, whereas retrieval-based methods retain
the full cache and select the relevant tokens with an indexer at every step. The
indexers differ in score function, selection granularity, and cadence
(Table~\ref{tab:op_diversity}). Training-free methods score pages with
representative keys~\cite{quest}, cluster the keys~\cite{starc, retroinfer}, or
hash them~\cite{magicpig}; MagicPIG and RetroInfer run the indexer on the CPU,
whose bandwidth and compute limit the $O(L)$ scan. Model-native designs train
the indexer with the model~\cite{deepseek-v3.2, glm-5.2, minimax-m3} and differ
from one another in all three axes. Further variants keep appearing, including
block attention trained into the model~\cite{moba, seer-atten}, switchable
dense--sparse attention~\cite{infllmv2}, and cross-layer index
reuse~\cite{indexcache, glm-5.2}, so new models keep changing the operations that a
serving system runs.

\subsection{Offloading the KV Cache}
LMCache~\cite{lmcache} and Mooncake~\cite{mooncake} place the KV cache in host
DRAM, whose capacity is bounded by the number of DIMM slots, and
HiSparse~\cite{hisparse} adds a hot buffer in HBM to reduce the resulting
transfers. CXL-attached memory expands capacity further at low
latency~\cite{pond, tpp}, and several systems place the KV cache
there~\cite{octopus, beluga, tract, sac}, while Strata~\cite{strata} extends the
hierarchy to storage and LIA~\cite{lia} moves model parameters instead. These
systems add capacity but leave the computation on the GPU, so the $O(L)$ index
scan and the gather of the selected entries cross a bandwidth- and
latency-limited link on every decoding step.

\subsection{Hardware Architecture for LLM Decode}
Since decode is memory-bound, many architectures place compute near memory.
AttAcc~\cite{attacc} offloads attention to HBM-PIM beside a GPU,
NeuPIMs~\cite{neupims} beside an NPU, and CHIME~\cite{chime} offloads it to
DIMM-PIM; NeuPIMs also interleaves sub-batches so that the two device types run
concurrently. Duplex~\cite{duplex} raises the logic-die operational intensity to
serve GQA attention and MoE, and IANUS~\cite{ianus} and CENT~\cite{cent}
distribute the model across PIM and PNM devices. All are provisioned for the low
operational intensity of dense MHA and GQA, below the intensity of MLA and of
the retrieval indexer (Table~\ref{tab:models}). Those built on HBM or GDDR also
provide far less capacity per watt than LPDDR (\S\ref{sec:arch_device}), and PIM
designs give up part of the array to logic, reducing usable
capacity~\cite{samsung_hbm_pim}.

LPDDR-based PNM and PIM designs instead attach large capacity to enable larger
batches and longer contexts~\cite{cxl-pnm}. Meridian~\cite{meridian} uses
LPDDR-based PIM to decompose document attention for RAG, and
HybridSpec~\cite{hybridspec} splits the draft and target models across a
hybrid-bonding stack and LPDDR5X for speculative decoding.
M$^2$NDP~\cite{m2ndp} makes the compute general-purpose, but provisions
far less of it than the retrieval indexer requires.
Beyond LPDDR, Raptor~\cite{raptor} stacks logic on DRAM to stream the KV cache,
and SHyLA~\cite{shyla} stores it in a 3D-stacked NVM--DRAM hybrid.
Stratum~\cite{stratum} pairs GPUs with near-memory processing on monolithic 3D
DRAM and tiers expert weights by predicted usage, targeting the MoE layers
rather than the KV cache. Hybe~\cite{hybe} pairs GPUs with NPUs sized to
saturate their memory bandwidth, running prefill on the GPUs and decode on the
NPUs.

Some designs run the token selection in the memory device, each specialized to a
single algorithm.
SPNM designs select token pages in a CXL
device~\cite{cxl-llm-pact25}, STARC~\cite{starc} selects clusters with PIM, and
LongSight~\cite{longsight} selects by dot-product top-$k$ on a compute-enabled
CXL expander~\cite{drex}. HILOS~\cite{hilos} runs attention on near-storage
accelerators, but targets offline inference.
None of these provides compute that is both general-purpose
and sized to the intensity of a retrieval indexer. \gpnm{} places
general-purpose compute next to large LPDDR5X capacity, sized to the operational
intensity of the operations it runs, so the indexers that models define execute
where the KV cache and index keys reside.

\section{Conclusion}
\label{sec:conclusion}

In this work, we presented a heterogeneous decode-phase serving system for
million-token contexts in LLMs that use retrieval-based sparse attention. Its
GPU nodes hold the model weights and run the projections and MoE layers, while
\gpnm{} nodes hold the KV cache and index keys and run every operation that
reads them. A \gpnm{} device combines large LPDDR capacity with general-purpose
compute sized for the retrieval indexer. To reduce pipeline bubbles between the
two device types, Opportunistic, Fine-grained Micro-batch Scheduling (OFMS)
hides the expert all-to-all behind the GEMMs of the other micro-batch.
Context-length-aware Micro-batch Rebalancing (CMR) then equalizes the token
counts of the two micro-batches despite the variance in context length. Across
three state-of-the-art models and real agentic traces, our system improves
throughput per TDP by $2.09$--$6.13\times$ over a GPU-only baseline, and it runs
training-free sparse attention methods without hardware change, improving each
by $1.36$--$3.21\times$. Because the \gpnm{} compute is not tied to one
algorithm, the same device can support the sparse attention mechanisms of later
model generations.

\bibliographystyle{IEEEtranS}
\bibliography{refs}

\end{document}